\documentclass[twocolumn,showpacs,preprintnumbers,amsmath,amssymb,pra]{revtex4}

\usepackage{graphicx}

\def\>{\rangle}
\def\<{\langle}
\def\ave#1{\left\langle #1\right\rangle}
\def\ii{{\rm i}}

\begin{document}

\title{Stability of quantum Fourier transformation on Ising quantum computer}

\author{Giuseppe Luca Celardo}
\affiliation{Dipartimento di
Matematica e Fisica, Universit\`a Cattolica, via Musei 41, 25121
Brescia, Italy}
\email{celardo@dmf.bs.unicatt.it}
\author{Carlos Pineda}
\affiliation{Universidad Nacional Aut\'onoma de M\'exico, Apdo.
Postal 20-364, Mexico D. F. 01000, Mexico}
\email{carlosp@fisica.unam.mx}
\author{Marko \v{Z}nidari\v{c}}
\affiliation{
Physics Department, Faculty of Mathematics and Physics, 
University of Ljubljana, Slovenia
}
\email{znidaricm@fiz.uni-lj.si}

\date{\today}

\begin{abstract}
We analyze the influence of errors on the implementation of the quantum 
Fourier transformation (QFT) on the Ising quantum computer (IQC). Two kinds 
of errors are studied: (i) due to spurious transitions caused by pulses and 
(ii) due to external perturbation. The scaling of errors with system 
parameters and number of qubits is explained. We use two different procedures 
to fight each of them. To suppress spurious transitions we use correcting pulses 
(generalized $2\pi k$ method) while to suppress errors due to external perturbation 
we use an improved QFT algorithm. As a result, the fidelity 
of quantum computation is increased by several orders of magnitude and is thus 
stable in a much wider range of physical parameters.
\end{abstract}


\maketitle

\section{Introduction}

Quantum information theory~\cite{Bennett} is a rapidly evolving field. 
It uses quantum systems to process information and by doing so can achieve 
things that are not possible with classical resources. Quantum secure communication for 
instance is already commercially available. Quantum computation on the other 
hand is still 
far from being useful outside of the academic community. Two serious 
obstacles to overcome in building quantum computers are: (i) one must 
be able to control the evolution in order to precisely implement quantum 
gates, (ii) one must suppress all external influences. Errors in both cases 
are caused by the perturbation of an ideal quantum computer, either due to 
the internal imperfections in the first case or due to coupling with the 
``environment'' in the second case. In the present paper we study both kind 
of errors in QFT algorithm and try to 
minimize them.
\par
In order to be closer to the experimental situation we choose a 
concrete model of a quantum computer, namely 
the Ising quantum 
computer (IQC)~\cite{Ising1st}. 
IQC is one of the simplest models still having enough complexity
to allow universal quantum computation.
Quantum gates on this computer can be realized 
by the application of electromagnetic pulses. For the algorithm we choose 
to discuss QFT. The first reason to choose QFT is that it is one of 
the most useful quantum algorithms, giving exponential speedup over the best 
classical procedure known, and is also one of the ingredients of some other 
important algorithms, e.g. Shor's factoring algorithm~\cite{Shor}. The 
second reason is that it is a 
complex algorithm, where by complex we mean it has more than ${\cal O}(n)$ number 
of quantum gates as opposed to previously studied more simple algorithms 
where the number of gates scales only linearly with the size of the computer 
(e.g. entanglement protocol~\cite{DynFid}). Previous study of Shor's algorithm in 
IQC~\cite{IsingShor} did not use recently introduced 
generalized $2 \pi k$ method~\cite{carlos} which is the best known procedure 
for inducing transitions on IQC. It is easy to 
imagine that in most useful quantum algorithms the size of the 
program will grow faster than linearly with the number of qubits $n$ and 
therefore it is important to see how errors accumulate in such algorithms. 
This importance is confirmed by our results showing that errors due to unwanted 
transitions for QFT grow with the \textit{square} of the number of pulses 
and not \textit{linearly} as in algorithms with linear ${\cal O}(n)$ number 
of gates, for a typical state.
\par
For QFT algorithm running on IQC we analyze errors 
due to spurious transitions caused by pulses, these we call 
\textit{intrinsic errors}, and errors due to the coupling with an external 
``environment'', called \textit{external errors}, modeled by a random hermitian 
matrix from a gaussian unitary ensemble (GUE)~\cite{Guhr}. We minimize intrinsic errors by 
applying some additional pulses to correct most probable errors~\cite{carlos} 
and by doing this, we are able to suppress intrinsic errors by several orders 
of magnitude. To suppress external errors due 
to GUE perturbation we use previously proposed improved quantum Fourier 
transformation (IQFT)~\cite{IQFT} which is more stable against GUE 
perturbations in a certain range of parameters. By using correlation 
function approach~\cite{Corr} we analyze in detail the dependence of 
errors on all relevant parameters and on 
the number of qubits $n$. By doing this we can set the limits between which 
parameters of IQC should lay in order to preserve the 
stability of computation. In our approach to decrease errors we do not use 
error correcting codes for the following reasons: we want to remove as many 
errors as we can on the lowest possible level and second, the intrinsic and 
external errors are not easily handled by error correcting 
codes (see Ref.\cite{Andrew} and references therein).
\par
The outline of the paper is as follows. In section II we repeat the
definition of IQC 
and in section III we summarize linear response formalism which
is the main theoretical tool for studying 
fidelity.
In section IV we study intrinsic and external errors, 
first separately and then the case of both errors present at the same time. 
In the appendix we present the pulse sequences used to implement QFT and IQFT 
algorithms.

\section{Ising quantum computer}

IQC consists of a 1-dimensional chain
of $n$ equally spaced identical spin $1/2$ particles
coupled by nearest neighbor Ising interaction of strength $J$,
so that parallel spins are favored over anti-parallel ones
by an energy difference of $J$ (we set $\hbar=1$ throughout
the paper). The quantum computer is operated via 
an external magnetic field having two components.
The first one is a permanent magnetic field oriented in the $z$
direction with a constant gradient which allows for the selective
excitation of individual spins, while the second one is a 
sequence of $T$ circular polarized fields in the
$x$-$y$ plane (which are called pulses) with different frequencies $\nu^{(m)}$, 
amplitudes (proportional to the Rabi frequencies $\Omega^{(m)}$),
phases $\varphi^{(m)}$ and durations $\tau^{(m)}$ for the $m$th pulse,
in which is encoded the protocol.
A particular orientation of the register 
allows to suppress
the dipole-dipole interaction between spins~\cite{magicangle1,magicangle2}.

The Hamiltonian of the system is
\begin{equation}\label{eq:hamiltonian}
\hat{H}=-\frac{1}{2}\sum_{l=0}^{n-1}\omega_l\hat{\sigma}_l^z
-\frac{J}{2}\sum_{l=0}^{n-2}\hat{\sigma}_l^z\hat{\sigma}_{l+1}^z
-\sum_{m=1}^{T} \hat{V}^{(m)}(t)\Theta^{(m)}(t)
\end{equation}
with
\begin{equation}
\hat{V}^{(m)}(t)=\frac{\Omega^{(m)}}{4}
\sum_{l=0}^{n-1}
(\hat{\sigma}_l^-\exp\{-\ii(\nu^{(m)} t+\varphi^{(m)}) \}+ \text{h.c.}),
\end{equation}
$\Theta^{(m)}(t)$ equal to one during the $m$th 
pulse and zero otherwise,
$\hat{\sigma}_l^{x,\, y, \, z}$ the usual Pauli operators
for spin $l$ and $\hat{\sigma}_l^\pm=\hat{\sigma}_l^x 
\pm \ii \hat{\sigma}_l^y$. Due to the constant gradient of the 
permanent magnetic field, the Larmor frequencies depend linearly on $l$,
 $\omega_l = (l+1)a$. By appropriately choosing the energy units we 
fix $J=1$ throughout the paper so that the only relevant energy 
scales are $\Omega^{(m)}$ 
and $a$. The basis states are chosen so that 
$\hat{\sigma}_l^{z} |0\>_l = |0\>_l$. 

We will introduce the following notation for further
discussion. Let the pulse
$P_i^{ac}$ indicate a pulse with frequency $\nu_i^{ac}$
resonant with the flip of the $i$-th spin if its neighbors
are in states ``$a$'' and ``$c$''. This will induce the
resonant transition $|\dots a_{i+1} b_i c_{i-1} \dots \rangle 
\to |\dots a_{i+1} \bar{b}_i c_{i-1} \dots \rangle $,  
named $T_i^{ac}$, if the pulse is a $\pi$ pulse 
($a,b,c \in \{0,1\}$). Note that for edge qubits, i.e. 
$i\in \{0,n-1\}$, only
one superscript is needed. 

Operating IQC in the \textit{selective excitation regime}, 
$\Omega^{(m)} \ll J \ll a$, allows one to separate transitions induced by 
pulses into three sets: \textit{resonant}, \textit{near-resonant} and 
\textit{non-resonant} according to the
detuning $\Delta$ of the transition which is the difference between the frequency
of the pulse and the energy difference of the states involved in the transition.
If $\Delta$ is exactly equal to zero, the transition is called resonant 
($T_i^{ac}$ induced by the pulse $P_i^{ac}$), 
if $\Delta$ is of the order of $J$ it is called
near-resonant 
($T_i^{a'c'}$ induced by the pulse $P_i^{ac}$ with 
$\{a',c'\}\ne\{a,c\}$), 
and if $\Delta$ is of the order of $a$ it is called
non-resonant
($T_{i'}^{a'c'}$ induced by the pulse $P_i^{ac}$ with 
$i'\ne i$).
In the implementation of a protocol resonant transitions are the
ones wanted, while near-resonant and non-resonant transitions are
a source of error. 

In the two level approximation~\cite{magicangle1}, a given unwanted 
transition with detuning $\Delta$ is induced with probability
\begin{equation}
p=\frac{\Omega^2}{\Omega^2+\Delta^2}\sin^2{\left(\rho \frac{\pi}{2}%
\sqrt{1+\frac{\Delta^2}{\Omega^2}} \right)},
\label{eq:p}
\end{equation}
where $\rho$ is a dimensionless duration of the pulse (for a $\pi$ pulse $\rho=1$ 
and for $\pi/2$ pulse it is $1/2$). The most probable transitions are the 
near-resonant ones and these can be suppressed 
as briefly described in the next paragraph.

For $P_i^{10}(=P_i^{01})$ pulses all near-resonant transitions have 
the same detuning so setting $\Omega$ to $\Delta/\sqrt{4k^2-1}$ with $k$ 
an integer suppresses these transitions.
Since for near-resonant transitions $\Delta={\cal O}(J)$, Rabi frequency
is for all pulses of the order of $\Omega \approx J/k$.
On the other hand, 
for $P_i^{00}$ and $P_i^{11}$ pulses near-resonant transitions have two 
different detunings therefore it is impossible to suppress both with a single pulse.
This problem can be overcome adding an additional correcting $P_i^{10}$ pulse.
The combination of these pulses in order to suppress all near-resonant
transitions is called $Q$-pulse denoted by $Q^{ac}_{i\rho}$ when 
doing a $\rho \pi$ rotation of the $i\text{th}$ qubit
if neighbors are in states ``$a$'' and ``$c$''.
This method to eliminate near-resonant transitions
is called generalized $2\pi k$ method. We refer the interested
reader to Ref.~\cite{carlos} for further details.
$Q$-pulses are the basic building blocks of gates, which in turn are the 
building blocks of algorithms such as QFT and IQFT.

QFT for $n=4$ qubits can be written as
\begin{equation}
U_\text{QFT}={\text T} {\text A}_0 {\text B}_{01} {\text B}_{02} 
{\text B}_{03}{\text A}_1 {\text B}_{12} {\text B}_{13} {\text A}_2 {\text B}_{23} {\text A}_3.
\label{eq:QFT4}
\end{equation}
There are in total $n$ Hadamard ${\text A}$ gates , $n(n-1)/2$ 
two-qubit ${\text B}$ gates, ${\text B}_{jk}=\text{diag}\{ 1,1,1,\exp{(\ii \theta_{jk})} \}$, 
with $\theta_{jk}=\pi/2^{k-j}$ and one transposition gate ${\text T}$ which 
reverses the order of qubits (e.g. ${\text T}|001\>= |100\>$). In total 
there are $n(n+1)/2+1$ gates. 
IQFT algorithm~\cite{IQFT} for $n=4$ qubits is given by
\begin{eqnarray}
U_\text{IQFT}&=&\text{T}\text{A}_{0}
\text{R}_{01}\text{R}_{02}
\text{R}_{03}\text{G}_{01} \text{G}_{02}\text{G}_{03} \nonumber \\
&\ & \hspace{-1cm} \times \text{A}_{1}\text{R}_{12}\text{R}_{13}\text{G}_{12}%
\text{G}_{13}\text{A}_{2}\text{R}_{23}\text{G}_{23}\text{A}_3,
\label{eq:IQFT4}
\end{eqnarray}
where ${\text G}_{ij}:={\text R}^\dagger_{ij} {\text B}_{ij}$.
The ${\text R}$ gate is defined by
${\text R}_{ij}|\dots a_i \dots b_j \dots \> :=
(-1)^{b_j} |\dots a_i \dots (\overline{a_i} \oplus  b_j) \dots \>$. 
In total there are $n^2+1$ gates in IQFT, i.e. roughly two times 
as many as for QFT. 

Recall that 
the implementation of quantum gates on IQC is easier 
in the interaction frame. Therefore, pulse sequences used in the paper 
implement the intended gates in the interaction frame. 
Each gate for QFT or IQFT (Eqs.~(\ref{eq:QFT4}) and (\ref{eq:IQFT4})) 
must in turn be implemented by several pulses (see appendix). The number of 
pulses for QFT 
grows as $\sim 18n^3$ whereas it grows as $\sim 54n^3$ for IQFT. Note 
that this number can become very large, e.g. for IQFT and $n=10$ one has 
$44541$ pulses. 
Throughout the paper our basic unit of time will be either a gate 
(as written for instance in Eqs.~(\ref{eq:QFT4}) and (\ref{eq:IQFT4})) or a pulse. 
A single exception will be the paragraph discussing correlation function 
of intrinsic errors, where the basic unit is a $Q$-pulse, which is 
composed of one or two pulses. The reason is that $Q$-pulses are the 
smallest near-resonant corrected unit of generalized $2\pi k$ method. 

\section{Linear Response theory}
\label{sec:LR}
As a criteria for stability we will use the fidelity $F(t)$, defined as an 
overlap between a state $\psi(t)$ obtained by an evolution with an ideal 
algorithm and the perturbed $\psi_\delta(t)$ obtained by the perturbed 
evolution:
\begin{equation}
F(t)=|\ave{\psi_\delta(t) | \psi(t)}|^2,
\label{eq:F}
\end{equation}
where $|\psi(t)\>=U(t)|\psi(0)\>$ and $|\psi_\delta(t)\>=U_\delta(t) |\psi(0)\>$. 
To simplify matters we will assume time $t$ to be a discrete integer variable, 
denoting some basic time unit of an algorithm, like a gate or a pulse. The 
quantity measuring the success of the whole algorithm is the fidelity $F(t)$ 
at $t=T$ where $T$ denotes the number of gates (pulses). One of the most 
useful approaches to studying fidelity is using linear response 
formalism in terms of correlation function of the perturbation, for a review 
see Ref.~\cite{Pregledni}. This approach has several advantages. First it rewrites 
the complicated quantity fidelity in terms of a simpler one, namely the 
correlation function, simplifying the understanding of the fidelity. Second, 
the scaling of errors with the perturbation strength, Planck's constant 
and with the number of qubits is easily deduced. Furthermore, as in 
practice one is usually interested in the regime of high fidelity, linear 
response is enough. 
\par
First we will shortly repeat linear response formulas as they will 
be useful for our discussion later. Let us write an ideal algorithm 
up to gate $t$ as $U(t)$
\begin{equation}
U(t)=U_t U_{t-1}\ldots U_1,
\label{eq:U}
\end{equation}
where $U_i$ is the $i$-th gate (pulse). If $t=T$ we have a decomposition of a whole algorithm.
\par
The perturbed algorithm can be similarly decomposed into gates
\begin{equation}
U^\delta(t)=U^\delta_t U^\delta_{t-1}\ldots U^\delta_1.
\label{eq:Ud}
\end{equation}
Each perturbed gate $U^\delta_j$ is now written as 
\begin{equation}
U^\delta_j=\exp{(-\ii \delta V_j )} U_j,
\label{eq:V}
\end{equation}
where $V_j$ is the perturbation of $j$-th gate and $\delta$ is a 
dimensionless perturbation strength. For any perturbed gate one can 
find a perturbation generator $V$, such that the relation~(\ref{eq:V}) 
will hold. Observe that the distinction into perturbation strength 
$\delta$ and the perturbation generator $V$ in Eq.~(\ref{eq:V}) 
is somehow arbitrary. If one is given an ideal gate $U$ and a 
perturbed one $U^\delta$ one is able to calculate only a product 
$\delta V$. This arbitrariness can always be fixed by demanding for 
instance that the second moment of the perturbation $V$ in a given 
state equals to $1$, $\< V^2\>-\< V \>^2=1$.\par
To the lowest order in the perturbation strength fidelity can be written 
as~\cite{Corr}
\begin{equation}
F(t)=1-\delta^2 \sum_{t_1,t_2=1}^t C(t_1,t_2),
\label{eq:LR}
\end{equation}
where the correlation function of the perturbation is
\begin{equation}
C(t_1,t_2)=\<V_{t_1}(t_1) V_{t_2}(t_2) \> -\< V_{t_1}(t_1) \> \< V_{t_2}(t_2) \> 
\label{eq:C}
\end{equation}
with $V_j(t)= U^\dagger(t) V_j U(t)$ being the perturbation of $j$-th 
gate propagated by an ideal algorithm up to time $t$, i.e. in the 
Heisenberg picture. The brackets $\< \cdot \>$ denote the expectation
value in the initial state. Throughout the paper we use random gaussian 
initial states and average over many of them to reduce statistical 
fluctuations. 
 Note that the time dependence of the correlation 
function~(\ref{eq:C}) is due to two reasons: one is time dependence 
due to the Heisenberg picture (time index in brackets) and the second 
one is due to the time dependence of the perturbation itself 
(time index as a subscript), i.e. one has 
different perturbations $V_j \neq V_k$ for different gates $j,k$. 
The expression for 
the fidelity Eq.~(\ref{eq:LR}) is the main result of the linear 
response theory of the fidelity. From this one can see that 
decreasing the correlation sum (or even making it zero, see Ref.~\cite{Freeze}) 
will increase the fidelity.
In Ref.~\cite{IQFT} stability of QFT algorithm was considered 
with respect to static 
GUE perturbation. Analyzing the correlation function
they were able to design an improved QFT algorithm (IQFT) 
which increases fidelity. 
\par
We are mainly interested in the fidelity $F(T)$ at the end of an algorithm. 
The final time $T$ in useful quantum algorithms depends on the number of 
qubits in a polynomial way, say as $T \propto n^p$. The power $p$ depends 
on the algorithm considered and of course also on our decomposition of an 
algorithm into gates (pulses). For QFT and IQFT algorithms with decomposition 
into gates, Eqs.~(\ref{eq:QFT4}) and (\ref{eq:IQFT4}), 
one has $p=2$. On the other hand, 
for the implementation of QFT on IQC one needs 
$T \propto n^3$ ($p=3$) basic electromagnetic pulses, as one is not able 
to directly perform ${\text B}_{jk}$ gates on distant qubits but has to 
instead use a number of pulses proportional to the distance between the 
qubits $|j-k|$. Now if the correlation function decays sufficiently fast, 
the fidelity will decay like $F=1-\delta^2 \sigma n^p$ whereas in the case 
of slow correlation decay the fidelity will decay as $F=1-\delta^2 c n^{2p}$. 
In the extreme case of perturbations at different gates being statistically 
uncorrelated (very fast decay of correlations) $\< V_j V_k \>\propto \delta_{jk}$ 
one obtains the exact formula $F=\exp{(-\delta^2 n^p)}$~\cite{IQFT}. In 
the limit of large quantum computer (large $n$) strongly correlated static 
errors, giving slow decay of correlations, 
will therefore be dominant due to fast $\propto n^{2p}$ growth. 
When we will discuss errors caused by perturbations due to the coupling with 
the environment we will focus on static perturbations, meaning the same 
perturbation on all gates, $V_k=V_j=V$, as this component will dominate 
large $n$ behavior.

\section{Errors in QFT}

Errors in an experimental implementation of QFT algorithm on 
an IQC can be of three kinds: (i) due to unwanted 
transitions caused by electromagnetic pulses (ii) due to coupling 
with external degrees of freedom and (iii) due to variation of system 
parameters in the course of algorithm execution. In the present paper 
we will discuss only the first two errors. 
Errors due to electromagnetic pulses are inherent to all algorithms 
on an IQC as we are presently unable to design 
pulse sequences for quantum gates without generating some unwanted 
transitions albeit with small probabilities. This errors can be in 
principle decreased by going sufficiently deep into selective excitation 
regime but one must of course keep in mind the limitations of real 
experiments\footnote{A new method for dealing with intrinsic
errors have been proposed recently in Ref.~\cite{lidar}}. 
Coupling with the ``environmental'' degrees of freedom is endemic in 
all implementations of quantum computers. As the environment will usually have 
many degrees of freedom we will model its influence on the quantum computer 
by some effective perturbation $V_\text{eff}$ given by a random matrix from a 
Gaussian unitary ensemble (GUE)~\cite{Guhr}. 
Note that coupling with the environment will generally cause
non-unitary evolution of the central system. We expect quantum computation to be
stable only on a time scale 
where evolution is approximately unitary,
i.e. for times smaller than the non-unitarity
time scale.
Therefore we limit ourselves to \textit{unitary} external perturbations.
The third kind of errors due to 
the variation of system parameters, like e.g. variation of Larmor 
frequencies due to the variation of the magnetic field is not considered 
in this paper. This does not mean they are not important.
Let us consider a systematic error in the 
gradient of the
magnetic field throughout the protocol ($a \to a+\delta a$). 
Demanding that the error in the largest eigenphase at the end
of the algorithm is much smaller than $1$, one gets the condition
$a/\delta a \approx a n^{p+2}/\Omega$. 
If one is in the selective excitation regime, this ratio can become
very large and this puts a severe demand on the experiments. 

To ease up understanding we will first 
discuss intrinsic errors only, then we will discuss external errors only 
and finally we will combine both errors.   

\subsection{Intrinsic Errors}
Let us consider probabilities of non and near-resonant transitions.
For near-resonant and non-resonant transitions we have 
$\Delta \gg \Omega$ and the probability given by the perturbation theory
is $p\propto (\Omega/\Delta)^2$.
In the $2\pi k$ method Rabi frequency $\Omega$ is given 
as $\Omega \sim J/k$ so the probabilities for near-resonant 
and non-resonant transitions are
\begin{eqnarray}
p^\text{near} &\propto & \left(\frac{1}{k} \right)^2 \nonumber \\ 
p^\text{non}_{jl} &\propto & \left( \frac{J}{ka(j-l)}\right)^2,
\label{eq:nearnon}
\end{eqnarray}
where $p^\text{non}_{jl}$ denotes probability of a non-resonant transition with 
$\Delta \approx a|j-l|$ involving $j$-th and $l$-th spin, one of which is a 
resonant one. The dependence of near and non-resonant errors on system
parameters is therefore different.
\par
For pulse sequences used to generate QFT or IQFT we always used a generalized 
$2\pi k$ method by which one can get 
rid of all near-resonant transitions. Therefore the only errors that remain are 
non-resonant ones. 
We first checked numerically 
that this is indeed the case by studying dependence of errors on system 
parameters by which one is able to distinguish near and non-resonant 
errors, Eq.~(\ref{eq:nearnon}).  

\begin{figure}[ht]
\includegraphics{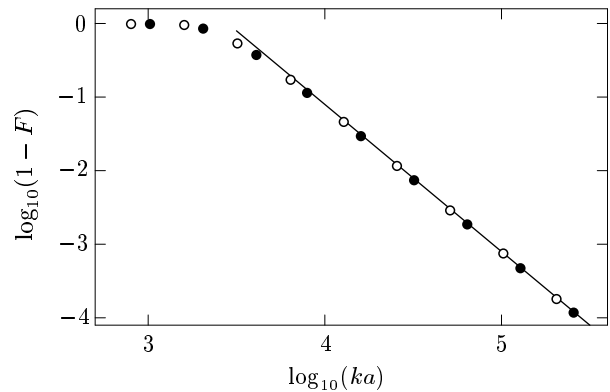}
\caption{\label{fig:presek}%
Dependence of fidelity on physical parameters of the system.
Empty points indicate variation of $k$ with $a=100$, filled points
indicate variation of $a$ with $k=128$, both for $n=6$.
Full line is theoretical dependence of $p^\text{non}$ given by 
Eq.~(\ref{eq:nearnon}).}
\end{figure}
As one can observe from Fig.~\ref{fig:presek} the agreement with the 
theoretical $p^\text{non}$ (Eq.~(\ref{eq:nearnon})) is excellent thereby confirming 
that the only errors left are non-resonant ones. By using generalized 
$2\pi k$ method we therefore decreased intrinsic errors by a factor of 
$(a/J)^2$ as compared to ordinary $2\pi k$ method where there are still some 
near-resonant errors present. 
In order to have complete understanding 
of fidelity decay due to intrinsic errors we have to understand scaling of 
these 
with the number of qubits. 
As we already discussed in 
section~\ref{sec:LR} this depends on two things: how strong the errors are 
correlated, giving possible scalings from $n^p$ to $n^{2p}$ and on 
the increase of the perturbation strength 
with the number of qubits. Let us 
first discuss the later. 
Under the assumption that the average transition probability 
(i.e. perturbation strength) for a non-resonant transition 
is the sum of all possible non-resonant transitions averaged over
all possible resonant qubits, we can estimate
\begin{equation}
\delta \propto \frac{1}{n} \sum_{j\neq l=0}^{n-1}{p^\text{non}_{jl}}\ \ 
\underrightarrow{n \to \infty} \ \ \left[ \frac{J}{ka}\right]^2 
\left( \frac{\pi^2}{3}-\alpha \frac{\log{n}}{n} \right),
\label{eq:avgp}
\end{equation}
with $\alpha$ some $n$ independent constant. We can see that the perturbation 
strength does not grow with $n$ asymptotically, but the convergence to 
its limit
is logarithmically slow. For small $n$ the perturbation 
strength therefore will grow with $n$ whereas it will saturate for large $n$. 
The second contribution to the $n$-dependence of the fidelity comes from the 
dynamical correlations between errors given by the correlation function~(\ref{eq:C}) 
of the perturbation generator for non-resonant 
errors. We numerically calculated this 
correlation function in order to understand how the correlation sum and therefore 
fidelity, Eq.~(\ref{eq:LR}), behaves as a function of $n$.

\begin{figure}[ht]
\includegraphics{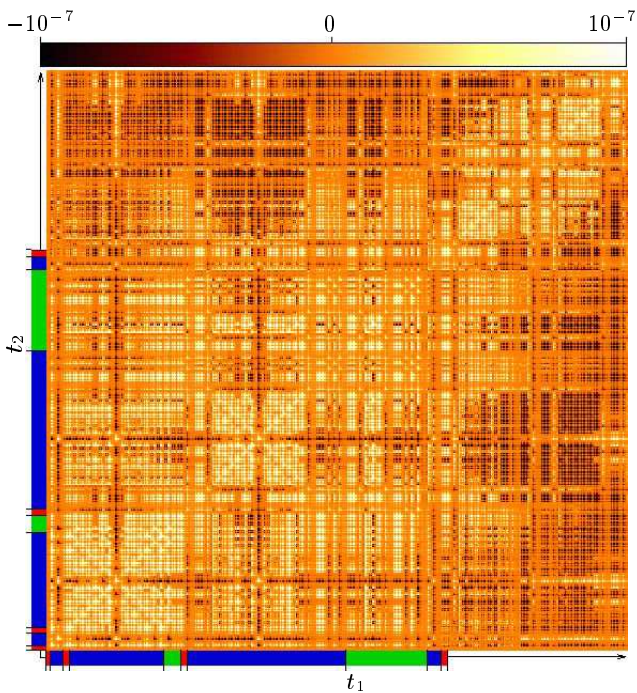}
\caption{\label{fig:corrfun} 
(Color online) Correlation function 
for intrinsic errors in QFT for $k=128$, $a=100$, $n=4$.
The shading on time axes denotes the duration of different 
gates, Eq.~(\ref{eq:QFT4}), and the time going from $1$ to $543$ runs over 
all $Q$-pulses.}
\end{figure}
\begin{figure}[ht]
\centering 
\includegraphics{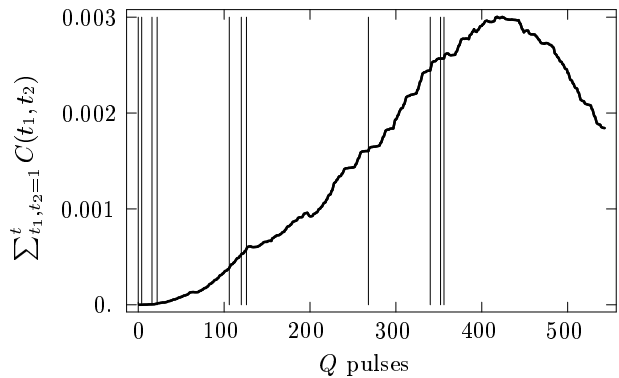}
\caption{\label{fig:corrsum}
The correlation sum of the same data as in 
Fig.~\ref{fig:corrfun}. Note 
the decrease of the sum when the transposition gate is applied. 
The fidelity is in this linear response regime simply given by
Eq.~(\ref{eq:LR}).
Vertical lines indicate beginning of each gate (Eq.~(\ref{eq:QFT4})).}
\end{figure}

\begin{figure}[ht]
\includegraphics{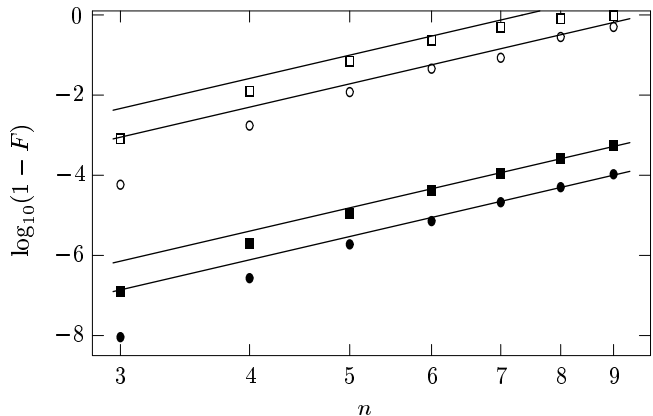}
\caption{\label{fig:nonN} 
Dependence of fidelity on the number of qubits.
Empty symbols indicate data for $k=128$ and $a=100$ while filled symbols
are for $k=1028$ and $a=1000$. 
Circles indicate QFT and squares IQFT. In the presence of
\textit{only} intrinsic errors, IQFT does not improve fidelity.
Full lines show asymptotic $n^6$ dependence.
}
\end{figure}

In Fig.~\ref{fig:corrfun} we show $C(t_1,t_2)$ averaged over all 
Hilbert space.
One can see that there are large 
2-dimensional 
regions of high correlations in all parts of the picture.
This means there are strong correlations between 
errors at different pulses and therefore the correlation sum will likely 
grow as $\sim n^{6}$ as the number of pulses scales as $n^3$
for our implementation of QFT. Similar results 
are obtained also for IQFT as can be inferred from Fig.~\ref{fig:nonN}.
One interesting thing to note 
is that during the application of the transposition gate at the end 
of the protocol the correlation sum starts to decrease at some point, 
nicely seen in Fig.~\ref{fig:corrsum} and also visible in the correlation 
picture in Fig.~\ref{fig:corrfun} as there are more negative than positive 
areas towards the end of the algorithm. This very interesting phenomena means 
that applying transposition at the end is advantageous (as compared 
to doing it classically for instance) as it will decrease non-resonant errors.
We checked that this principle can not be exploited further by repeating 
transposition many times and by this decreasing correlation sum even more. 
Still, this surprising behavior suggests that it might be possible
to decrease non-resonant errors in a systematic way.

To furthermore confirm predicted $\sim n^6$ growth of the correlation sum we
calculated the dependence of intrinsic errors on $n$. 
This can be seen in Fig.~\ref{fig:nonN}, where we plot $1-F(T)$ 
as a function of $n$ for QFT and IQFT and for two different sets of 
parameters, one for $k=128$, $a=100$ giving large errors and one for 
$k=1024$, $a=1000$. One can see that asymptotically for large $n$ the 
dependence is indeed $n^6$ but the convergence to this behavior is 
fairly slow, one needs of the order of $n=7$ or more qubits. This slow 
convergence we believe is due to the logarithmic convergence of the 
perturbation strength (Eq.~(\ref{eq:avgp})). To get exact coefficients 
in front of $n^6$ dependence we fitted dependences of errors in 
Fig.~\ref{fig:nonN} with a polynomial in $n$ using 
at most two nonzero terms. 
Defining polynomials in the linear response regime 
as $s^\text{in}=(1-F) (ka/J)^2$ one gets for QFT and IQFT 
\begin{eqnarray}
s^\text{in}_\text{QFT}(n) &=&280 n^6-660 n^5 \nonumber \\
s^\text{in}_\text{IQFT}(n) &=&1300 n^6-2100 n^5.
\label{eq:errIntrin}
\end{eqnarray}
Both expressions are good for $n \ge 5$ and superscript ``$\text{in}$'' 
denotes intrinsic errors. 
Beyond the linear response the exponential dependence is frequently 
justified~\cite{Corr} and one has
\begin{equation}
F=\exp{\left(-\left[\frac{J}{ka}\right]^2s^\text{in}(n)\right)}.
\label{eq:Fintrin}
\end{equation}
Large coefficients of
polynomials in Eqs.~(\ref{eq:errIntrin}) are due to large number of pulses. 
The maximum possible dependence in the case of no decay of correlation 
function (see discussion at the end of section~\ref{sec:LR}) could 
be $T^2$ and therefore the 
leading terms in polynomials~(\ref{eq:errIntrin}) 
expressed by the number of pulses are $s^\text{in}_\text{QFT}\sim 0.8 T_\text{QFT}^2$
and $s^\text{in}_\text{IQFT} \sim 0.5 T_\text{IQFT}^2$. Therefore relative to 
the number of pulses IQFT slightly decreases non-resonant errors but 
in the absolute sense QFT is better simply because it has only one third 
as many pulses as IQFT and the coefficient 
in front of $n^6$ (Eq.~(\ref{eq:errIntrin})) 
is thereby smaller. If only intrinsic errors in the 
generalized $2\pi k$ method are 
concerned QFT is always more stable than IQFT. 
Note that the intrinsic errors due to non-resonant transitions for QFT grow 
as $\sim T^2$ ($\sim n^6$) whereas in previously studied ``simple'' algorithms,
for instance entanglement protocol~\cite{DynFid}, they grow only as the first 
power of the number of gates $\sim T$. This means that QFT is much more 
sensitive to intrinsic errors.

\subsection{External Errors}
In order to study only external errors we set throughout this section
parameters to $k=1024$ and $a=1000$, for which intrinsic errors are
much smaller than external ones. 

External error will be modeled by the perturbation 
$V$ (Eq.~(\ref{eq:V})) chosen to be a random hermitian matrix from a GUE ensemble. 
To facilitate comparison with previous results on IQFT~\cite{IQFT} we will 
make perturbation after each quantum gate, except for the last 
transposition gate ${\text T}$, Eq.~(\ref{eq:QFT4}), 
after which we do not make perturbation. So for QFT we make $n(n+1)/2$ 
perturbations, while for IQFT we make $n^2$ perturbations. One other possible 
choice would be to make perturbations after each pulse. We will discuss 
this possibility at the end of this section. For now let us just say that 
qualitatively the results are the same as if doing perturbation after each 
gate, one just has to rescale perturbation strength like 
$\delta_\text{gate} \propto n \delta_\text{pulse}$ as there are effectively ${\cal O}(n)$ 
perturbations (pulses) per gate. 
\par
\begin{figure}[ht]
\includegraphics{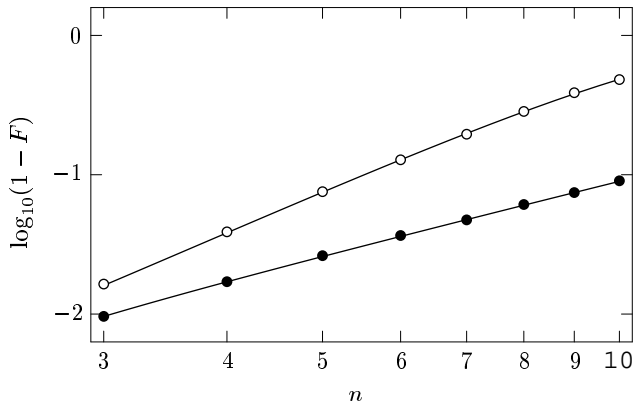}
\caption{\label{fig:Lgue} 
Dependence of $1-F$ in QFT algorithm on the number 
of qubits for static perturbation 
in the interaction frame (empty points) and in the laboratory frame 
(full points), with $\delta=0.04$. 
Lines are best fitting polynomials (see Eq.~(\ref{eq:sGueL})).}
\end{figure}

The implementation of QFT on IQC is written in the 
interaction picture. As the static perturbation is the worst, meaning 
it will asymptotically in large $n$ limit be dominant, we will concentrate 
only on static perturbation, i.e. the same perturbation for all gates (pulses) 
$V_j=V_k=V$. There are still two possibilities, either making static 
perturbation in the interaction frame or making it static in the laboratory 
frame. Let us first discuss the later case. If we make static perturbation 
in the laboratory frame, we can of course transform it to the interaction 
frame by a unitary transformation $W(t)=\exp(-\ii H_0 t)$ given by the
time independent part $H_0$ of the Hamiltonian Eq.~(\ref{eq:hamiltonian}). 
This transformation
\begin{equation}
\exp{(-\ii \delta V_\text{int}(t))}:=W^\dagger(t) 
\exp{(-\ii \delta V_\text{lab})} W(t),
\label{eq:Vlab}
\end{equation}
will result in the perturbation in the interaction frame $V_\text{int}(t)$ 
being time dependent. As the transformation to the laboratory frame $W(t)$ 
in the selective excitation regime involves large phases, perturbations 
at different gates will tend to be uncorrelated due to averaging out of 
widely oscillating factors in the correlation function~(\ref{eq:C}). 
Therefore in the first approximation one can assume 
$C(t_1,t_2) = \delta_{t_1,t_2}$ and the fidelity will in this extreme 
case of uncorrelated errors decay as $F(T)=\exp{(-\delta^2 T)}$~\cite{IQFT}. 
On the other hand, if we make perturbation to be static in the interaction 
frame and the correlation function is not Kronecker delta in time, the 
fidelity will decay faster with the number of qubits (or $T$). 
To numerically confirm these arguments, we show in Fig.~\ref{fig:Lgue} 
dependence of fidelity with the number of qubits $n$ for the two cases discussed, 
static perturbation in the interaction frame and static perturbation in the 
laboratory frame.
Polynomial fitting of $n$ dependence for QFT gives
\begin{eqnarray}
s^\text{gue}_\text{lab}(n)&=&0.47 n^2+1.41 n-2.42 \nonumber \\
s^\text{gue}_\text{int}(n)&=&0.45 n^3-0.42 n^2+0.58 n.
\label{eq:sGueL}
\end{eqnarray}
The fidelity due to external GUE errors is given as
\begin{equation}
F=\exp{(-\delta^2 s^\text{gue}(n))},
\label{eq:Fgue}
\end{equation}
with the appropriate $s^\text{gue}(n)$ from Eq.~(\ref{eq:sGueL}). Note that 
$s^\text{gue}_\text{int}(n)$ grows faster than $s^\text{gue}_\text{lab}(n)$ 
as argued. Observe also that for the static perturbation in 
the laboratory frame using the assumption of uncorrelated errors in the 
interaction frame we predicted $s^\text{gue}_\text{lab} \approx T \approx n^2/2$ 
for QFT which is remarkably close to the numerically observed value 
Eq.~(\ref{eq:sGueL}). For IQFT and the application of GUE perturbation 
in the laboratory frame one gets a similar result with the leading term 
$s^\text{gue}_\text{lab}(n) \sim 1.12 n^2$.
One can write an arbitrary time dependent perturbation in the interaction 
frame as a Fourier series and for large $n$ the static component will always prevail. 
Therefore, 
from now on we will exclusively discuss only static perturbations in the 
interaction frame. 
\par
\begin{figure}[ht]
\includegraphics{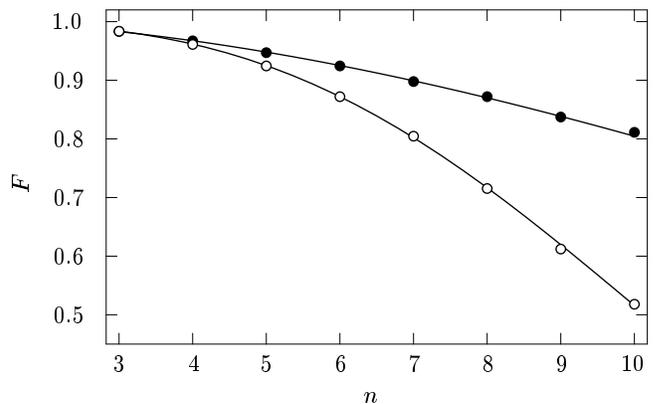}
\caption{\label{fig:gue} 
Dependence of fidelity on the number of qubits for 
QFT (empty symbols) and IQFT (filled symbols) algorithms ($\delta=0.04$). 
Curves are theoretical prediction Eq.~(\ref{eq:Fgue}) using polynomials 
from Eqs.~(\ref{eq:sIQFT}) and (\ref{eq:sGueL}).}
\end{figure}

The dependence of errors due to GUE perturbation in the case of QFT and IQFT 
has already been derived~\footnote{Taking into account 
different definition of fidelity in Ref.~\cite{IQFT},
polynomials are almost the same with the slight 
difference due to the different number of applied perturbations.}.
For IQFT numerical fitting in our case gives dependence
\begin{equation}
s^\text{gue}_\text{IQFT}(n)=1.31 n^2+0.86 n-3.73.
\label{eq:sIQFT}
\end{equation}
Here the perturbation is again static in the interaction frame 
as is the case throughout the paper with a single exception being 
the previous paragraph. Dependence of fidelity in both cases for QFT 
and IQFT can be seen in Fig.~\ref{fig:gue}, together with the theory
(Eq.~(\ref{eq:Fgue})) using polynomials (\ref{eq:sIQFT}) and (\ref{eq:sGueL}).

Observe that IQFT for $n > n_\text{crit}=3$ is better than QFT 
despite having more gates and therefore applying perturbation on it 
more times (for $n\le 3$ QFT is slightly better). What is important 
is that the dependence of errors on $n$ is also different, $\sim n^3$ 
for QFT, but only $\sim n^2$ for IQFT. This means that asymptotically 
IQFT is much more stable against GUE perturbations than ordinary QFT.
\par
\begin{figure}[ht]
\includegraphics{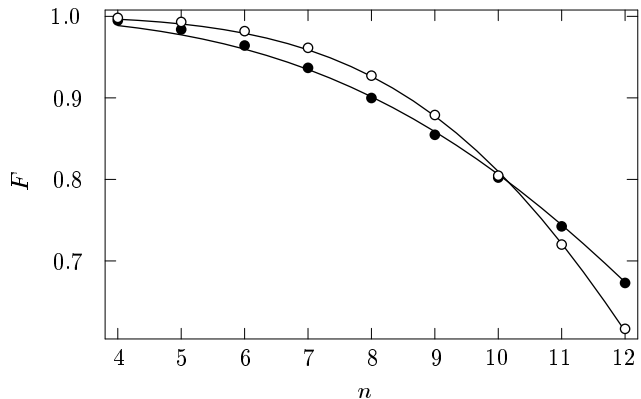}
\caption{\label{fig:each} 
Dependence of $F$ on the number of qubits for the static GUE 
perturbation after each pulse with $\delta=5 \cdot 10^{-4}$. 
Empty symbols are for QFT and filled symbols are for IQFT.
Curves are theoretical 
prediction Eq.~(\ref{eq:Fgue}) using best fitting polynomials given 
by Eq.~(\ref{eq:sEach}).}
\end{figure}

Finally, let us discuss what happens if we make static GUE perturbations 
in the interaction frame after \textit{each pulse}, and not after each gate
as done so far. 
The product of two operators 
$\exp{(-\ii \delta V)} U$ can be written 
as $\exp{(-\ii \delta V)}U = U \exp{(-\ii \delta V(1))}$, with 
$V(1):=U^\dagger V U$. Using this expression for all errors 
within a single gate and bringing them all to the beginning of the gate we get
\begin{eqnarray}
\exp{(-\ii \delta V)}U_r  
\cdots \exp{(-\ii \delta V)}U_1 \approx \nonumber \\
U_r \cdots U_1 \exp{(-\ii \delta [ V(1)+\cdots+V(r)] )},
\label{eq:commute}
\end{eqnarray}   
where $r$ is number of pulses 
constituting a gate.
This means that the application of the perturbation after each pulse is 
to the lowest order in $\delta$ \textit{equivalent} to the application of the effective 
perturbation $\delta \sum_j^r{V(j)}$ after the gate. 
Of course now the perturbation 
is explicitly time dependent. 
But for a GUE matrix acting on a whole 
Hilbert space individual pulses will do  transformations on an
exponentially small subspace (i.e. on one qubit)
and therefore one might expect 
that effectively one can write $\delta \sum_j^r{V(j)} \approx r V_\text{eff}$, 
where $V_\text{eff}$ is some effective random matrix independent of the gate. 
As in our case of QFT on IQC we have on average $\propto n$ pulses for 
a single gate we can predict that doing perturbation with strength 
$\delta_\text{pulse}$ after each pulse is approximately equal as doing 
perturbation of strength $\delta_\text{gate} \approx n \delta_\text{pulse}$ 
after each gate. In order to confirm these expectations we did numerical 
experiments with the results shown in Fig.~\ref{fig:each}. 
Fitting polynomial in the dependence of fidelity, Eq.~(\ref{eq:Fgue}), 
for QFT and IQFT gives in this case
\begin{eqnarray}
s^\text{gue}_\text{QFT}(n)&=&4.86 n^5+35.8 n^4 \nonumber \\
s^\text{gue}_\text{IQFT}(n)&=&25.6 n^4+606 n^3.
\label{eq:sEach}
\end{eqnarray}
The leading dependence of $n^5$ for QFT and $n^4$ for IQFT nicely 
agrees with our rescaling prediction $\delta_\text{gate} \approx n \delta_\text{pulse}$. 
IQFT is asymptotically again better than QFT as the errors grow slower 
with the number of qubits. The crossing point between the two in this case 
happens at $n_\text{crit}=10$, whereas in the case of perturbation after each gate 
we had $n_\text{crit}=3$. This confirms that doing GUE perturbation after each 
pulse is qualitatively the same as doing it after each gate, only the crossing
point between QFT and IQFT changes and of course also the dependence of errors on $n$ 
changes, simply due to the different number 
of applied perturbations. If perturbation strength 
$\delta$ is properly rescaled, the $n$ dependence is the same in both cases. 
\par
Up to now we discussed intrinsic errors and external errors separately. 
The next
question is of course, what happens if both errors are present at the same time 
and are of similar strength?

\subsection{Intrinsic and External Errors combined}
If both kinds of errors are present, a first naive 
guess would be that they just add,
\begin{equation}
F^\text{both}=F^\text{in} F^\text{gue}=
\exp{\left(-\left[\frac{J}{ka}\right]^2 s^\text{in}(n)-\delta^2 s^\text{gue}(n) \right)},
\label{eq:Fboth}
\end{equation}
with the appropriate polynomials $s^\text{in}(n)$ and $s^\text{gue}(n)$
given in previous Eqs.~(\ref{eq:errIntrin}), 
(\ref{eq:sGueL}) and (\ref{eq:sIQFT}). 
In the linear response regime this formula means that both
errors are uncorrelated, i.e. their cross-correlations are zero. This is 
easy to proof using properties of GUE matrices. Let us calculate 
cross-correlation function between $V^\text{in}(t_1)$ and $V^\text{gue}(t_2)$ 
averaged over GUE ensemble. Written explicitly one has to average products 
of the form $V^\text{in}_{ij} V^\text{gue}_{jk}$, where $V^\text{gue}$ is a 
GUE matrix. As this expression is linear in $V^\text{gue}$
it averages to zero, $\< V^\text{in}_{ij} V^\text{gue}_{jk} \>_\text{gue}=0$, 
thereby explicitly confirming a simple additions of both errors. Of course 
in real experiments we are not averaging over GUE ensemble but are taking 
one definite representative member of it. But for large Hilbert 
space the expectation value of a typical random state and one 
particular GUE matrix is ``self-averaging'' and will be equal to the ensemble 
average. 
\par
\begin{figure}
\includegraphics{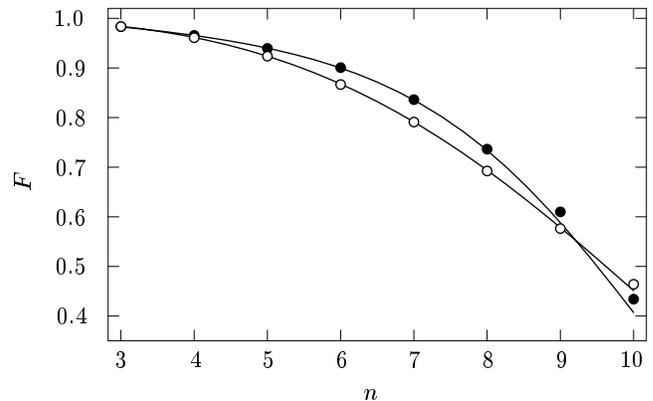}
\caption{\label{fig:both} 
Fidelity for QFT (empty symbols) and IQFT (filled symbols) algorithm 
and GUE perturbation after 
each gate. System parameters are $k=a=200$ and $\delta=0.04$ (intrinsic
and external errors are comparable in size). Full curves are theoretical predictions 
for $F$ given by Eq.~(\ref{eq:Fboth}).} \end{figure}

Let us check the theoretical prediction for fidelity Eq.~(\ref{eq:Fboth}) 
with a numerical experiment. We again apply GUE perturbation after each gate.
The results together with the theoretical prediction Eq.~(\ref{eq:Fboth}) 
are in Fig.~\ref{fig:both}. The agreement between the theory and the experiment 
is good also beyond the linear response regime. Please note that we 
deliberately choose parameters so that both QFT and IQFT give similar 
fidelity in order to also see the crossing of the two curves within
the shown range of $n$.
Given fixed $\delta$ and $ka$, QFT is always better for large $n$
because intrinsic errors will prevail over external ones, 
due to their fast $n^6$ growth.
But still, for intermediate $n$'s IQFT can be better that QFT as seen in 
Fig.~\ref{fig:both}.
\par
Now we are equipped with understanding of errors in QFT and IQFT 
due to external GUE perturbation and intrinsic errors so
we can make some predictions regarding ranges of experimental 
parameters $ka$, $\delta$, $n$ for which the fidelity will be high enough. 
Interesting question for instance is, when is 
IQFT better than QFT? To find that, we set $F_\text{QFT}=F_\text{IQFT}$ with 
$F$'s given by Eq.~(\ref{eq:Fboth}). This results in the condition
\begin{equation}
\delta_\text{crit}=\frac{J}{ka}\sqrt{\frac{s^\text{in}_\text{QFT}-
s^\text{in}_\text{IQFT}}{s^\text{gue}_\text{QFT}-s^\text{gue}_\text{IQFT}}}.
\label{eq:dcrit}
\end{equation}
For $\delta> \delta_\text{crit}$ IQFT is better than QFT.
\begin{figure}[ht]
\includegraphics[height=3.3in,angle=-90]{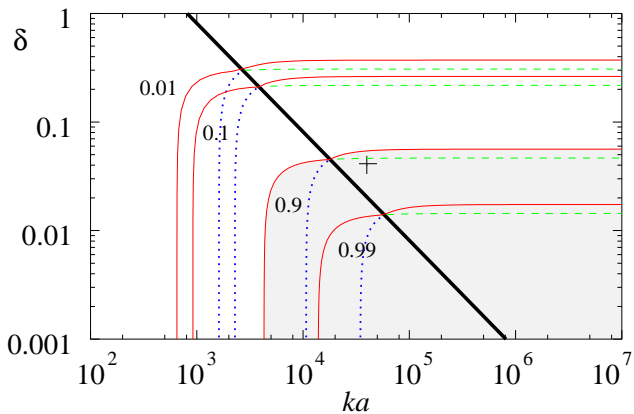}
\caption{\label{fig:dqn5} 
(Color online)~Dependence of fidelity on system parameter $ka$ 
and GUE perturbation strength $\delta$ for $n=5$. 
Full curves of constant fidelity are composed of two parts corresponding 
to QFT or IQFT. Above the thick line for $\delta_\text{crit}$ 
IQFT is better and below QFT is 
better. Dotted curves of constant fidelity below this line are for IQFT 
and dashed lines above are for QFT. The shaded region corresponds to the
 region of fidelity larger than $0.9$. The plus symbol shows the position 
of parameters for Fig.~\ref{fig:both}.}
\end{figure}
In Fig.~\ref{fig:dqn5}
we show curves of constant fidelity for $n=5$. They are 
composed of two parts, above the line for $\delta_\text{crit}$ IQFT is 
better than QFT, and below vice versa. Two characteristic features are also 
vertical and horizontal asymptotes of the curves of constant fidelity. 
The vertical asymptote means that 
for fixed $n$, even if $\delta=0$, we must have $ka$ larger than some
critical value 
determined just by intrinsic errors, in order to have given fidelity. 
Horizontal asymptote for high $ka$ means that if $\delta$ is larger than 
some critical value, increasing $ka$ will not help to improve fidelity.
\begin{figure}[ht]
\includegraphics[height=3.3in,angle=-90]{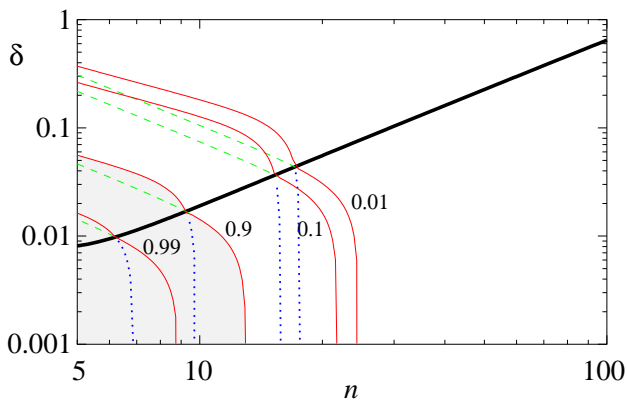}
\caption{\label{fig:crossQ} 
(Color online)~Fidelity dependence on $\delta$ and number of 
qubits $n$ for a fixed value of $ka=10^5$. For the explanation 
of various curves see the caption to Fig.~\ref{fig:dqn5}.}
\end{figure}
\begin{figure}[ht]
\includegraphics[height=3.3in,angle=-90]{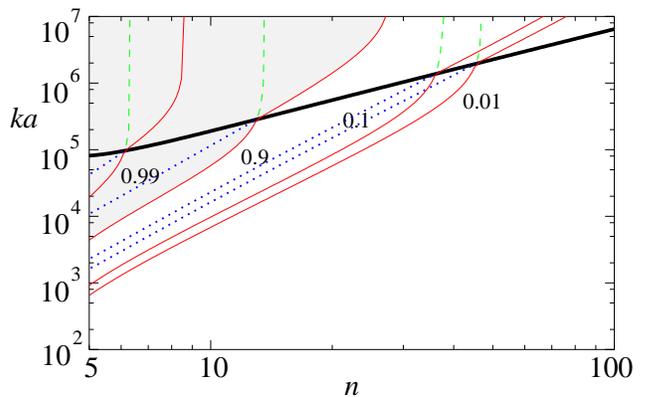}
\caption{\label{fig:crossD} 
(Color online)~Fidelity dependence on $ka$ and $n$ for a 
fixed $\delta=0.01$. For the explanation of various curves see the 
caption to Fig.~\ref{fig:dqn5}. 
}
\end{figure}
In Figs.~\ref{fig:crossQ}~and~\ref{fig:crossD} we show similar plots, 
only these time one of the axes is dependence on $n$. For instance, from 
Fig.~\ref{fig:crossQ} on can see that having $ka=10^5$, 
the maximum number of qubits is $n \approx 12$ if we want to have fidelity 
larger than $0.9$ (even if $\delta=0$). 
This unfavorable growth of 
required $ka \propto n^3$ in order to have a fixed fidelity
is due to $\sim n^6$ growth of intrinsic errors. It would 
therefore be advantageous to find a way to suppress errors due to
non-resonant transitions~\cite{kamenev}.
 
\section{Conclusions}
We analyzed two possible errors in the implementation of QFT on IQC 
working in the selective excitation regime. 
We consider:
(i) intrinsic errors due to unwanted transitions caused by pulses, 
(ii) external errors due to 
coupling with the external degrees of freedom. 
We carefully analyze
their dependence on system parameters and on the number of qubits. To diminish
intrinsic errors we use the generalized $2\pi k$ method by which we are able to 
suppress all near-resonant transitions, with only much smaller non-resonant 
transitions remaining. We then study these non-resonant errors in QFT algorithm 
and by using correlation function formalism explain their growth with 
time as $\sim T^2$,
in contrast to so far studied ``simple'' algorithms (having ${\cal O}(n)$
gates), 
where the growth is linear in time.
The immediate question is whether this behavior is general for algorithms
having more than ${\cal O}(n)$ gates.
This very fast growth with $n$ is a consequence of strong 
correlations between errors at different pulses and puts a severe demand 
on experimental requirements. 
Therefore it would certainly be desirable to find a way to suppress
also non-resonant errors.
We also consider  
perturbations due to coupling with external degrees of freedom modeled by a 
random GUE matrix. To suppress this kind of errors we show that it is advantageous 
to use an improved QFT algorithm, for which the errors grow only as 
$\sim n^2$, whereas they grow as $\sim n^3$ for ordinary QFT. By a combination of 
both techniques, the generalized $2\pi k$ method and improved QFT algorithm, 
we are able to 
make implementation of QFT stable in a much wider range of parameters. 

\begin{acknowledgments}
Useful discussions with T.~Prosen, T.~H.~Seligman, G.~Berman, 
B.~Borgonovi and R.~Bonifacio, are gratefully 
acknowledged. The work of C.P. was supported by
Direcci\'on General de Estudios de Posgrado (DGEP).
C.P. is thankful to the University of Ljubljana and
the Universita Cattolica at Brescia
for hospitality. The work of M.~\v Z. has been financially supported 
by the Ministry of Science, Education and Sport of Slovenia.
\end{acknowledgments}

\appendix*

\section{QFT and IQFT implementation on the Ising Quantum Computer}

To implement the protocol with high fidelity we use
$Q_{i\rho}^{ab}$ pulses derived in Ref.~\cite{carlos},
which completely suppress all near-resonant errors.
Phases of $Q$-pulses composing a gate must be chosen correctly so that
the gate works on an arbitrary state.
The protocols implementing
$\text{CN}_{ij}$ (control not gate) and $\text{N}_{j}$ (not gate) 
can be found in sections 7.1-7.3 of Ref.~\cite{carlos}.

In order to complete QFT and IQFT we still need
to implement the $\text{R}^\dagger$, $\text{R}$, $\text{A}$, 
$\text{B}$ and $\text{T}$
gates. 
We can decompose $\text{R}$, $\text{R}^\dagger$ and 
$\text{T}$ gates into simpler pieces:
\begin{eqnarray}
\text{R}_{ij}&=& \text{N}_i \text{CN}_{ij}  \text{N}_i  \text{Z}_j,  \\*
\text{R}_{ij}^\dagger &=& \text{N}_i \text{Z}_j \text{CN}_{ij}  
\text{N}_i,\text{ and}  \\*
\label{eq:T}
\text{T}&=& \prod_{i=1}^{q} \prod_{j=1}^{q-i}\text{S}_{q-j,q-j-1}
\end{eqnarray}
with $\text{S}_{ij}=\text{CN}_{ij}\text{CN}_{ji}\text{CN}_{ij}$ the swap
gate, $\text{Z}=\text{diag}\{1,-1\}$ the $\sigma_z$ gate and
each term in the product in Eq.~(\ref{eq:T}) is placed at the left of the sub-product (e.g.
$\prod_{i=0}^2 D_i=D_2 D_1 D_0$).
Therefore, the only gates left to design are $\text{A}$, $\text{B}$ and
$\text{Z}$.

The phases of $Q$-pulses can be expressed in terms of angles
$\theta_\rho$, $\alpha_\rho$, 
$\Theta_\rho$, $\beta_\rho$ and $\gamma_\rho$~\cite{carlos}
which are given by
\begin{eqnarray}
\theta_\rho&=&\pi\sqrt{k_\rho^2-\rho^2/4},\\*
\alpha_\rho&=&\frac{\pi}{2}\sqrt{k_\rho^2+3\rho^2/4},\\*
\tan \Theta_\rho&=&-\frac{\theta_\rho}{2\alpha_\rho}\tan \alpha_\rho,\\*
\tan \beta_\rho&=&-\frac{\pi}{2\alpha_\rho}\tan \alpha_\rho \cos\Theta_\rho,\\*
\gamma_\rho&=&\sqrt{(\pi k_\rho)^2-(\pi+\beta_\rho)^2}.
\end{eqnarray}
We use 
notation of angles without subscripts
denoting angles for $\pi$ pulses
i.e. $\theta \equiv \theta_1$ and set $k_{1/2}=2k$.

The Hadamard gate can now be expressed as
\begin{eqnarray}\label{eq:Ain}
\text{A}_{j}&=&
Q_{j}^{00}(\varphi_1)
Q_{j}^{10}(\varphi_2)
Q_{j}^{11}(\varphi_3)
Q_{j\frac{1}{2}}^{00}(\varphi_4)
\nonumber \\& \ & 
Q_{j\frac{1}{2}}^{10}(\varphi_5)
Q_{j\frac{1}{2}}^{11}(\pi/2),
\end{eqnarray}
for intermediate qubits and
\begin{equation}\label{eq:Aedge}
\text{A}_{j} = 
Q_{j}^{0}(\varphi_6)
Q_{j}^{1}(\varphi_7)
Q_{j\frac{1}{2}}^{0}(\varphi_8)
Q_{j\frac{1}{2}}^{1}(\pi/2),
\end{equation}
for edge qubits, with
\begin{equation}
\begin{array}{rclrcl}\label{eq:Ainphases}
\varphi_1&=& -2 \left( \theta+\gamma_\frac{1}{2}+
\theta_\frac{1}{2} \right),
&\varphi_2&=&-\theta-2\Theta,
\\
\varphi_3&=& -2\left(\theta+\gamma-\gamma_\frac{1}{2}-
\theta_\frac{1}{2}\right),
&\varphi_4&=&\pi/2-2\gamma-4\theta_\frac{1}{2},
\\
\varphi_5&=& \pi/2-\theta_\frac{1}{2}-2\Theta_\frac{1}{2},
&\varphi_6&=&-\theta-\theta_\frac{1}{2},
\\
\varphi_7&=&-\theta+\theta_\frac{1}{2},
&\varphi_8&=&\pi/2-2\theta_\frac{1}{2}.
\end{array}
\end{equation}
For neighboring qubits ($|i-j|=1$) the $\text{B}$ gate 
can be written as,
\begin{eqnarray}\label{eq:Bin}
\text{B}_{ij}&=&
Q_{i}^{11}(0) 
Q_{i}^{10}(0) 
Q_{i}^{00}(0) 
Q_{j}^{10}(0) 
Q_{j}^{10}(\varphi_1)
Q_{j}^{00}(0)
\nonumber \\& \ & 
Q_{j}^{00}(\varphi_2) 
Q_{i}^{11}(\varphi_3)
Q_{i}^{10}(\varphi_3)
Q_{i}^{00}(\varphi_3) 
Q_{j}^{10}(0)
\nonumber \\& \ & 
Q_{j}^{10}(\varphi_4) 
Q_{j}^{11}(0)
Q_{j}^{11}(\varphi_5),
\end{eqnarray}
for intermediate qubits and for edge qubits ($i$ or $j\in\{0,n-1\}$)
it is
\begin{eqnarray}\label{eq:Bedge}
\text{B}_{ij}& = &
Q_{i}^{1}(0)      
Q_{i}^{0}(0)
Q_{j}^{10}(0)      
Q_{j}^{10}(0)      
Q_{j}^{00}(0)
\nonumber \\& \ & 
Q_{j}^{00}(\varphi_6)
Q_{i}^{1}(\varphi_7)
Q_{i}^{0}(\varphi_8)
Q_{j}^{10}(0)     
\nonumber \\& \ & 
Q_{j}^{10}(\varphi_9)
Q_{j}^{11}(0)      
Q_{j}^{11}(\varphi_{10}).
\end{eqnarray}
Angles for $\text{B}$ gates are
\begin{equation}
\begin{array}{rclrcl}\label{eq:Binphases}
\varphi_1&=&-2\gamma-3\theta+2\Theta,
&\varphi_2&=&\phi/2-2\gamma-6\theta,
\\
\varphi_3&=&\phi/4-\pi/2,
&\varphi_4&=&-\varphi_1,
\\
\varphi_5&=&\phi/2+2\gamma+6\theta,
&\varphi_6&=&\phi/2-6\gamma-12\theta+4\Theta,
\\
\varphi_7&=&\varphi_3-\varphi_1,
&\varphi_8&=&\varphi_3+\varphi_1,
\\
\varphi_9&=&-2\varphi_1,
&\varphi_{10}&=&\phi/2-2\gamma+4\Theta,
\end{array}
\end{equation}
and $\phi=\pi/2$.
For distant qubits ($|i-j|>1$) it is necessary to use swap gates to 
bring $i$-th and $j$-th qubits to neighboring positions,
then apply 
$\text{B}$ protocol for neighbor qubits
and finally take them back to their original
positions using swap gates. 
The angle $\phi$ in Eq.~(\ref{eq:Binphases}) is in this case 
$\phi=\pi/2^{|j-i|}$.
Finally the $\text{Z}$ gate is expressed as
\begin{equation}\label{eq:Ppin}
\text{Z}_{j}=
    Q_{j}^{11}(0)
    Q_{j}^{10}(0)
    Q_{j}^{00}(0)
    Q_{j}^{11}(\pi/2)
    Q_{j}^{10}(\pi/2)
    Q_{j}^{00}(\pi/2)
\end{equation}
for intermediate qubits and
\begin{equation}\label{eq:PpinE}
\text{Z}_{j}=
      Q_{j}^{1}(0)
      Q_{j}^{0}(0)
      Q_{j}^{1}(\pi/2)
      Q_{j}^{0}(\pi/2)
\end{equation}
for edge qubits.
Counting the number of all pulses for QFT and IQFT one gets
\begin{eqnarray}
T_{\text{QFT}}&=& 18n^3-16n^2-49n+57\\*
T_{\text{IQFT}}&=&54n^3-86n^2-105n+191. 
\end{eqnarray}
\label{appendix}


\end{document}